\documentclass[useAMS,usenatbib]{mn2e}
\usepackage{epsf}

\title[Modeling peculiar velocities of dark matter halos]{
Modeling peculiar velocities of dark matter halos\thanks{
This paper was published in Mon.~Not.~R.~Astron.~Soc.~{\bf 343}, 
1312-1318 (2003). 
Owing to an error in numerical computations, some incorrect results were 
presented there.
Erratum is to be published in Mon.~Not.~R.~Astron.~Soc. 
Conclusions of the original version are unaffected by the correction.
This version supersedes the original version.}}

\author[T. Hamana et al.]
{Takashi Hamana$^{1,2}$, Issha Kayo$^3$, Naoki Yoshida$^{2,4}$, Yasushi
Suto$^{3,5}$ and Y.P. Jing$^6$\\
$^1$ Institut d'Astrophysique de Paris, 98bis Boulevard Arago, 
F 75014 Paris, France\\ 
$^2$ National Astronomical Observatory of Japan, Mitaka, Tokyo 181-8588, 
Japan\\
$^3$ Department of Physics, University of Tokyo, Tokyo 113-0033, Japan\\
$^4$ Harvard-Smithsonian Center for Astrophysics, 60 Garden Street, 
Cambridge, MA 02138, USA\\
$^5$ Research Center for the Early Universe(RESCEU), School of Science, 
University of Tokyo, Tokyo 113-0033, Japan\\
$^6$ Shanghai Astronomical Observatory, the Partner Group of MPI f\"{u}r 
Astrophysik, Nandan Road 80, Shanghai 200030, China}

\date{Accepted 2003 May 5, Received 2003 April 16; 
in original form 2003 January 9}
\pagerange{\pageref{firstpage}--\pageref{lastpage}} 
\pubyear{2003} 

\begin{document}

\label{firstpage}

\maketitle

\begin{abstract}
We present a simple model that accurately describes various statistical
properties of peculiar velocities of dark matter halos.  We pay
particular attention to the following two effects.
First, the evolution
of the halo peculiar velocity depends on the local matter density,
instead of the global density.  Secondly, dark matter halos are biased
tracers of the underlying mass distribution, thus halos tend to be
located preferentially at high density regions.  For the former, we
develop an empirical model calibrated with $N$-body simulations, while
for the latter, we use a conventional halo bias model based on the 
extended Press-Schechter model combined with an
empirical log-normal probability distribution function of the mass
density distribution.  We find that, compared with linear theory, the
present model significantly improves the accuracy of predictions of
statistical properties of the halo peculiar velocity field including the
velocity dispersion, the probability distribution function, and the
pairwise velocity dispersion at large separations.  Thus, our model
predictions may be useful in analyzing future observations of the
peculiar velocities of galaxy clusters.
\end{abstract}

\begin{keywords}
galaxies: clusters: general --  cosmology: theory -- dark matter --
large-scale structure of universe 
\end{keywords}

\section{Introduction}

In the standard structure formation scenario, dark matter halos are
thought to be formed via the gravitational amplification of initial
small density fluctuations, and their motion results from gravitational
forces acting over large scales.  As a consequence, the
root-mean-square (rms) peculiar velocities of halos at the present time
depend on the total amount of matter in the universe, $\Omega_0$, and
on the normalization of the matter power spectrum, $\sigma_8$, as
$\propto\sigma_8 \Omega_0^{0.6}$.  In addition, linear theory predicts
that the halo peculiar velocity evolves as $\propto a(t) dD(t)/dt$ where
$a(t)$ is the scale factor normalized unity at the present time and
$D(t)$ is the linear growth rate (see Section 2).  Thus, a measurement of
the rms peculiar velocities of halos can be used as a test of structure
formation scenarios in principle.

A measurement of the peculiar velocity of a dark matter halo is possible
through its baryon contents.  Because a cluster of galaxies is formed in
the gravitational potential of a massive dark matter halo, its peculiar
motion traces that of the hosting halo.  Therefore, the rms peculiar
velocity of clusters of galaxies provides a reasonable measure of that
of halos.  So far, empirical distance indicators such as the
Tully--Fisher and the $D_n$--$\sigma$ relations have been used to measure
the peculiar velocity of relatively nearby systems (e.g. Bahcall,
Gramann \& Cen 1994; Lauer \& Postman 1994; Bahcall \& Oh 1996;
Moscardini et al. 1996; Borgani et al. 1997; 2000; Watkins 1997; Dale et
al. 1999; Hudson et al. 1999; Colless et al. 2001).  Future observations
of the kinematic Sunyaev--Zel'dovich effect will provide measurements of
line-of-sight peculiar velocity for distant systems (Sunyaev \&
Zel'dovich 1980; Rephaeli \& Lahav 1991; Haehnelt \& Tegmark 1996;
Kashlinsky \& Atrio-Barandela 2000; Aghanim, G\'orski \& Puget 2001).

Theoretical predictions of the rms peculiar velocity of halos have been
made by applying the peak theory (Bardeen et al. 1986), in which halo
motions are modeled by motions of density peaks of a suitably smoothed
version of the initial Gaussian density field evolved according to
linear theory (Peebles 1980).  $N$-body simulations have been used to
test those predictions (e.g. Bahcall, Gramann, Cen 1994; Croft \&
Efstathiou 1994; Suhhonenko \& Gramann 1999, Colberg et al.~2000; Sheth
\& Diaferio 2001; Yoshida, Sheth \& Diaferio 2001).  Colberg et
al. (2000) found that, at $z > 4$, the mean peculiar velocities of
materials that end up in massive halos agree well with the predictions
of linear theory for the corresponding peaks, but that linear theory
systematically underestimates their subsequent growth; the rms peculiar
velocities at $z=0$ evaluated from $N$-body simulations are larger than
the linear theory prediction by up to 40 percent.  This discrepancy may
be ascribed to the following facts: that evolution of peculiar velocities 
of halos
in high density regions are accelerated relative to the average (Tormen
\& Bertshinger 1996; Cole 1997; Kepner, Summers \& Strauss 1997) and
that halos tend to be formed in such higher-density regions because of
the biased structure formation mechanism.  Sheth \& Diaferio (2001)
have shown using $N$-body simulations that the evolution of the halo
peculiar velocity indeed depends on the local matter density, rather
than on the global density.  They presented a simple model for the shape
of the distribution function of the peculiar velocities that takes into
account their dependence on the local density in an empirical manner.

The purpose of this paper is to present a simple and accurate model for
the rms of peculiar velocities of halos.  Hereafter, we denote the rms
peculiar velocity of halos with mass $M$ as $\sigma_{\rm halo}(M)$.  

The major improvements that we achieve here are to take into account the
following two points that play an important role in understanding
statistical properties of halo peculiar velocities. (i) The evolution of
the halo peculiar velocity depends on the local matter density, rather
than on the global density.  We develop a model of the local 
density-dependent growth of the halo peculiar velocity 
in an empirical manner following Sheth \& Diaferio (2001). 
(ii) Dark matter halos are
biased tracers of the dark matter distribution, and thus halos tend to
be located at high-density regions.  In order to implement this, we use
the halo biasing model developed by Mo \& White (1986) and Sheth \& Tormen 
(1999; 2002) combined with an
empirical log-normal probability distribution function (PDF) of the mass
density distribution (Kayo, Taruya \& Suto; 2001).

The outline of this paper is as follows.  In Section 2, we summarize
linear theory predictions for the peculiar velocities of peaks. In
Section 3, we describe $N$-body simulations that are used to calibrate
and to test the model.  We construct the model in Section 4.  The model
predictions are compared with $N$-body simulations in section 5.  A
summary and discussion are presented in Section 6.

\section{Linear predictions for the peculiar velocities of peaks}

In this section, we summarize basic equations of the linear
theory predictions for peculiar velocities of density peaks.

According to linear theory of gravitational instability in a dust
universe (Peebles 1980), the peculiar velocity field of mass 
grows as
\begin{eqnarray}
\label{eq:linearv}
v &\propto& a \dot{D} 
= \dot{a} D  {{d \ln D} \over {d \ln a}} \nonumber\\
&=& \dot{a} D(a)\left[{{5 \Omega_0} \over {2 D(a) X(a)^2}}
+ {{\Lambda_0 a^2 - {{\Omega_0}/{2 a}}}\over {X(a)^2}} 
-1 \right],
\end{eqnarray}
where 
\begin{equation}
\label{eq:X}
X(a)={{\dot{a}} \over {{H_0}}}
=\left[{{\Omega_0} \over a}+(1-\Omega_0 -\Lambda_0)
+\Lambda_0 a^2 \right]^{1 \over 2}.
\end{equation}
The linear growth factor $D(a)$ is given by
\begin{equation}
\label{eq:D}
D(a)={{5 \Omega_0} \over 2} {{X(a)} \over a} 
\int_0^a {{da'} \over {X(a')^3}}.
\end{equation}
Note that the frequently used approximation $f(a)=d\ln D/d\ln a
\simeq \Omega^{0.6}$ underestimates the growth rate 
by $\sim 5$ percent in the flat $\Omega_0=0.3$ cosmology that we
consider here. Thus, we will not adopt the approximation but we will 
use the numerical integration throughout the present paper.

We assume that the primordial (linear) density fluctuations obey the
Gaussian statistics. Then the peculiar velocity field is isotropic and
Gaussian, and its three-dimensional dispersion smoothed over a smoothing
scale of $R$ is given by
\begin{equation}
\label{eq:sigmav}
\sigma_v(R,a)=\dot{a} f(a) \sigma_{-1}(R,a).
\end{equation}
Following Bardeen et al.~(1986), $\sigma_j$ is defined for any integer
$j$ as
\begin{equation}
\label{eq:sigmaj}
\sigma_j^2(R) = {1 \over {2\pi^2}} \int dk~ k^{2+2j} P(k) W^2(kR),
\end{equation}
where $W(x)$ is the Fourier transform of the smoothing window, and
$P(k)$ is the linear matter power spectrum.  Throughout this paper, we
use the real space top-hat window function which corresponds to
$W(x)=(3/x^3)[\sin(x)-x\cos(x)]$.  We adopt the fitting function of a
cold dark matter (CDM) power spectrum given by Bardeen et al.~(1986), 
who show that the rms peculiar velocity at the peaks of the smoothed density
field differs systematically from $\sigma_v$ by
\begin{equation}
\label{eq:sigmap}
\sigma_p(R)=\sigma_v(R) \sqrt{1-\sigma_0^4/\sigma_1^2\sigma_{-1}^2}.
\end{equation}
Note that this expression does not depend on the height of peaks.

The conventional assumptions in modeling the peculiar velocities of
halos are as follows. 
(i) The velocities of halos are identical to those of peaks
(equation \ref{eq:sigmap}).  The smoothing scale is chosen such that the
filtered mass is equal to that of the halo considered, i.e., $M=4\pi
\bar{\rho} R^3/3$ in the case of the real space top-hat window function.
(ii) The halo velocities evolve according to linear theory 
(equation~\ref{eq:linearv}).  However, the prediction based on these
assumptions seems to underestimate the rms peculiar velocities of halos
(Colberg et al. 2000).

\section{$N$-body simulation}

In constructing an improved model 
of the peculiar velocity of halos, we first use
$N$-body simulations of a $\Lambda$CDM cosmology ($\Omega_0=0.3$,
$\Lambda_0=0.7$, $h=0.7$, and $\sigma_8=1.0$).  The details of the
$N$-body simulations are described in Jing \& Suto (1998); see also 
Kayo et al.~(2001).  Briefly, the simulations employ $256^3$ CDM
particles in a cubic box of $300\,h^{-1}$Mpc on a side.  The simulations
are performed using a P$^3$M code with the gravitational softening
length of $\epsilon \sim 120\,h^{-1}{\rm kpc}$.  The initial matter
power spectrum is computed using the CDM fitting function given by
Bardeen et al.~(1986).  We have three realizations which differ only in
the phases of the initial density fluctuations.  The particle mass
($m_{\rm part}=1.34\times 10^{11}h^{-1}M_\odot$) of the simulations is
sufficiently small to guarantee that there are practically no
discreteness effects on dark matter clustering on scales down to the
softening length in the redshift range of interest for our purposes
(Hamana, Yoshida \& Suto 2002).

We identify dark matter halos using the standard friends-of-friends
algorithm with a linking parameter of $b=0.164$ (in units of the mean
particle separation).  We set the minimum mass of the halos as
$1.34\times 10^{12} h^{-1}M_{\odot}$, which corresponds to the mass of
10 simulation particles.  We define the peculiar velocity of each halo
to be the center-of-mass peculiar velocity of member particles.

The smoothed mass density fields are computed on $64^3$ regular grids in
the simulation box using the top-hat window function.  For each halo,
the smoothed mass density of the nearest grid is assigned as its local
background density.

\begin{figure}
\begin{center}
\begin{minipage}{8.4cm}
\epsfxsize=8.4cm 
\epsffile{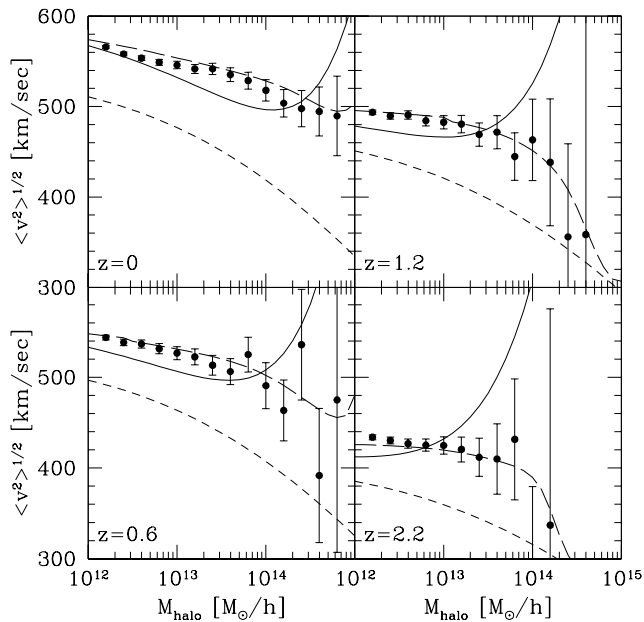}
\end{minipage}
\end{center}
\caption{Halo velocity dispersions as a function of halo mass at
different redshifts as indicated in each panel.  
Filled circles are measurements from $N$-body simulations.  
The long-dashed lines represent the phenomenological model, 
equation~(\ref{eq:sigma_halo_M}) adopting the conditional mass function via 
equation~(\ref{eq:pMdelta2}), while solid lines are for the model with the 
approximation equation (\ref{eq:pMdelta}).
The dashed lines show the linear theory 
prediction for peaks (equation~\ref{eq:sigmap}).}  
\label{fig:sigmav_z}
\end{figure}

Fig.~\ref{fig:sigmav_z} shows the rms peculiar velocities of halos in
the $N$-body simulation (filled circles) together with the linear theory
prediction of peaks, (equation~\ref{eq:sigmap}; dashed curves) as well as
our predictions described in the next section 
(solid lines and long-dashed lines).
Note that, in order to account
for the finite size of the simulation box ($L_{\rm box}=300h^{-1}$Mpc),
we set a lower limit of the integration in equation~(\ref{eq:sigmaj}) by
$k_{min}=2\pi/L_{\rm box}$.  This large-scale cut-off reduces 
$\sigma_p$ by a fractional amount of about 8 percent, and is applied to all
model predictions in this paper.
The plot confirms the previous finding that the linear theory predictions
underestimate the rms peculiar velocities by about 10--40 percent for cases
$0<z<2.2$ in a $\Lambda$CDM cosmology.  
In addition, it is seen from this Figure that the linear theory 
predictions have a slightly steeper slope than that of $N$-body
simulations, especially for the case of $z=0$.  This discrepancy may be
explained by the following: 
(i) massive halos tend to be formed in denser regions as a
result of the biasing mechanism, and (ii) halos in denser regions move
faster than those in less dense regions, as will be proved in the next 
section (Fig.~\ref{fig:sigmav_delta} and \ref{fig:condmf}). 
As a consequence, on average, deviations from
linear theory appear larger for massive halos, and thus the slope of
$N$-body data becomes flatter.

\section{Model}

In this section, we describe our model of the rms peculiar velocity of
halos.  We basically follow the framework developed by Sheth \& Diaferio
(2001); see also Diaferio \& Geller (1996) and Sheth (1996). They originally
developed the model for the distribution function of peculiar velocities
of dark matter particles and halos, but their framework can be
straightforwardly applied to the model for the rms of halo peculiar
velocities.

Sheth \& Diaferio (2001) pointed out that the evolution of the halo
peculiar velocity depends on the local matter density (defined with an
appropriate smoothing length which we discuss later).  We thus first
construct a model for the dependence of the halo peculiar velocity
dispersion, $\sigma_{\rm halo}^2(M,\delta)$, on the local background
density $\delta$.  Following Sheth \& Diaferio (2001), we consider a
parametric model
\begin{equation}
\label{eq:sigmahalo}
\sigma_{\rm halo}^2(M,\delta)=[1+\delta(R_{\rm local})]^{2\mu(R_{\rm local})} 
\sigma_p^2(M),
\end{equation}
where $R_{\rm local}$ is the smoothing scale with which the local
background density $\delta$ is defined.  The key question is how to
define the {\it appropriate} smoothing length-scale.  Evidently, $R_{\rm
local}$ should enclose the gravitational coherence scale which is
responsible for the local deviation of the peculiar motion of a halo
from its global value (for instance, given by linear theory).  
Thus, the size of halo (e.g., the virial radius) sets a lower 
limit for $R_{\rm local}$.  
On the other hand, choosing values for $R_{\rm local}$ that are too large
would result in an excessively smoothed density field, and local
density fluctuations, by which the deviation from the global value is
induced, are smoothed out.  We may, thus, expect that $R_{\rm local}$ is
in the range between the linear scale (such that $\sigma_0(R)<1$) and
the non-linear scale ($\sigma_0(R)>1$). 
Concerning the power-law index $\mu$, the linear theory relation, 
$f\propto \Omega^{0.6}$, suggests $\mu\simeq 0.6$ (Sheth \& Diaferio 2001).
Note, however, that $\mu$ depends on $R_{\rm local}$ on which the local 
density is defined (a larger $\mu$ for a larger smoothing length, because 
the range of $\delta$ becomes narrower), thus $\mu$ does not 
need to be very close $\mu\simeq 0.6$, but it is expected that with 
an appropriate choice of the smoothing length-scale, $\mu$ would be 
close to 0.6.

\begin{figure}
\begin{center}
\begin{minipage}{8.4cm}
\epsfxsize=8.4cm 
\epsffile{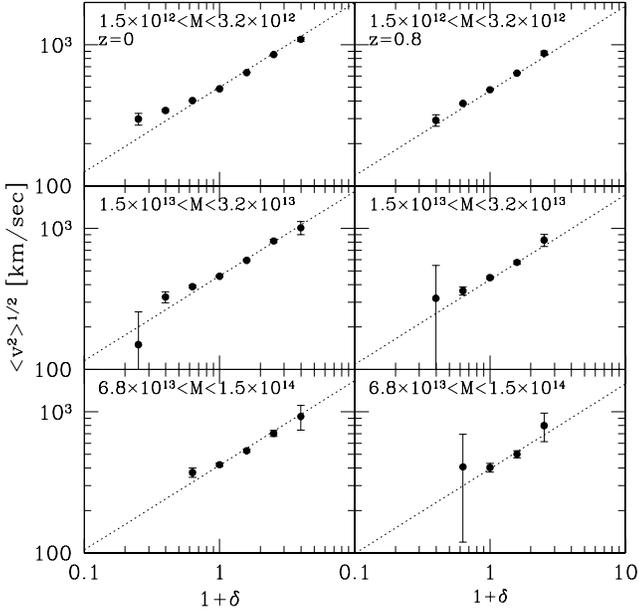}
\end{minipage}
\end{center}
\caption{Halo peculiar velocity dispersions as a function of the local
background density for limited mass ranges.  Halo mass ranges (in
units of $h^{-1}M_\odot$) are denoted in each plot.  
Dotted lines represent the model equation~(\ref{eq:sigmahalo}) with 
$\mu=0.6$ and filled circles show measurements from our $N$-body simulations.  
The smoothed local density fields are computed from the $N$-body particle 
distributions using the top-hat window function with a smoothing scale 
of $20h^{-1}$Mpc.}  
\label{fig:sigmav_delta}
\end{figure}

\begin{figure}
\begin{center}
\begin{minipage}{5cm}
\epsfxsize=5cm 
\epsffile{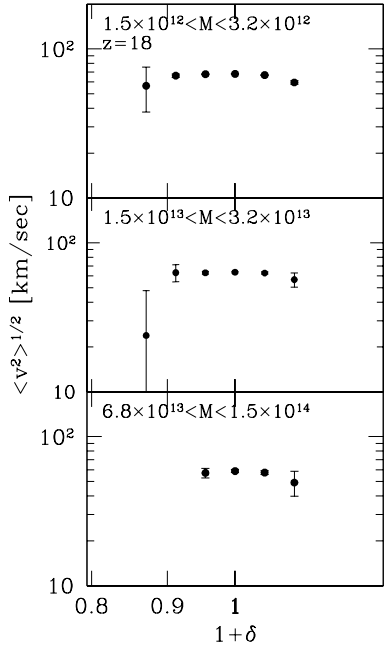}
\end{minipage}
\end{center}
\caption{Same as Fig.~\ref{fig:sigmav_delta} but for halo progenitors 
at the initial redshift of the $N$-body simulation ($z=18$). 
We mark the member particles of a halo identified at $z=0$,
and then compute the position and velocity of the halo progenitor from 
the center-of-mass position and velocity of the marked particles at $z=18$.
The top-hat
smoothing scales for computing the local density are taken to be 
$20h^{-1}$Mpc.}  \label{fig:sigmav_deltaini}
\end{figure}

To proceed further, we adopt an ansatz that $R_{\rm local}$ is
given by a relation $\sigma_0(R_{\rm local})=\sigma_{\rm local}$, and we
determine the model parameter $\sigma_{\rm local}$ empirically using
$N$-body simulations.  As expected from the discussion above, a
suitable choice of $\sigma_{\rm local}$ would be within a range
$0.1<\sigma_{\rm local}<1$.  We find that choosing $\mu=0.6$ (constant)
with $R_{\rm local}$ given by $\sigma_{\rm local}=0.5(1+z)^{-0.5}$ (through
$\sigma_0(R_{\rm local})=\sigma_{\rm local}$) provides reasonable fits
to the results of the $N$-body simulations as shown in 
Fig.~\ref{fig:sigmav_delta}.  
For this choice, $R_{\rm local}\simeq 18 $ and
$16h^{-1}$Mpc for z=0 and 0.8, respectively.  
The value of $\mu=0.6$ is consistent with
one suggested in linear theory ($\mu\simeq 0.6$). Thus, while
the choice of the smoothing length may still be arbitrary, we
decide to use the above sets of parameters for definiteness.

Fig.~\ref{fig:sigmav_deltaini} shows the peculiar velocity
dispersion of halo progenitors as a function of the local density at the
initial redshift of the $N$-body simulation $z=18$.  
We mark the member particles of a halo identified at $z=0$, and then 
compute the position and velocity of the halo progenitor from the 
center-of-mass position and velocity of the marked particles at $z=18$.
Fig.~\ref{fig:sigmav_deltaini}
clearly indicates that there is no clear correlation between the
peculiar velocity and the local density at the initial epoch.  
Together with Fig.~\ref{fig:sigmav_delta}, this strongly supports 
the idea that
most of the halo peculiar velocity is acquired through the subsequent
evolution of the nearby density field.  A closer look at 
Fig.~\ref{fig:sigmav_deltaini} reveals that the initial peculiar velocity
dispersions decline very slightly at both the lowest and highest bins.
An interpretation of this is that, at the initial epoch, the halo 
progenitors locate close to positive (negative) density peaks tend 
to have relatively higher (lower) local densities. 
At the same time, a potential gradient at positions close to the density peaks 
tends to be small, and thus the progenitors at such regions move slower 
than others.
As a result, the highest and lowest density bins have a slightly lower 
peculiar velocity than other bins.

As shown by Sheth \& Diaferio (2001), if the range of halo masses and
local background density is sufficiently small, the PDF of halo peculiar 
velocities is well
approximated by a Maxwellian distribution, i.e., the PDF of 
one-dimensional velocity is a Gaussian.  We denote by $p(M|\delta)$ the
probability of finding a halo with mass $M$ at a region where the
background density is $\delta$.  Then the peculiar velocity dispersion
of halos with mass $M$ is given by summing up the dispersion
$\sigma_{\rm halo}^2(M,\delta)$ weighted over the probability of finding
halos in regions of $\delta$:
\begin{equation}
\label{eq:sigma_halo_M}
\sigma_{\rm halo}^2 (M)=
{{\int d \delta~p(M|\delta) \sigma_{\rm halo}^2(M,\delta)} 
\over 
{\int d \delta~p(M|\delta)}}. 
\end{equation}
Note that this expression is valid only if the background density is
defined appropriately (i.e., with the smoothing scale $R_{\rm local}$
defined above).  The expression for $p(M|\delta)$ is obtained as
follows.

\begin{figure}
\begin{center}
\begin{minipage}{8.4cm}
\epsfxsize=8.4cm 
\epsffile{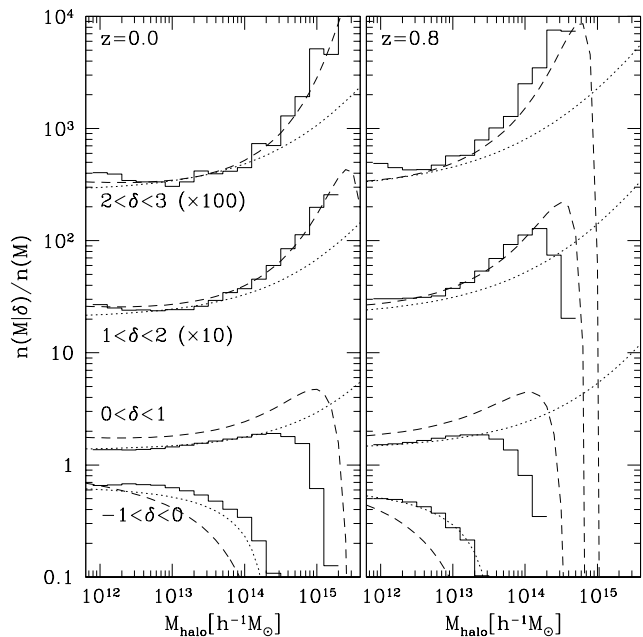}
\end{minipage}
\end{center}
\caption{Conditional mass functions normalized by the unconditional
mass function for ranges of the local background density, from lower 
to upper, $-1<\delta<0$, $0<\delta<1$, 
$1<\delta<2$ and $2<\delta<3$.
Dashed curves show the conditional mass function by Sheth \& Tormen (2002),
dotted curves show the approximation $n(M|\delta)\simeq [1+b(M)\delta] n(M)$
with the unconditional mass function and halo bias model by Sheth \& Tormen 
(1999), and the solid histograms are for $N$-body data.
The top two cases are offset upward by a factor of 10 and 100.
The smoothing scales for computing the local density are 
$R_{\rm local}=20$ and $10h^{-1}$Mpc 
for $z=0$ (left panel), and 0.8 (right panel), respectively.}  
\label{fig:condmf}
\end{figure}

It is well established that dark matter halos are biased tracers of the
underlying dark matter distribution.  
In this picture, it is useful to write the conditional probability as
\begin{equation}
\label{eq:pMdelta2}
p(M|\delta)={{n(M|\delta)}\over{n(M)}} p_{dm}(\delta),
\end{equation}
where $n(M)$ and $n(M|\delta)$ denote the halo mass function and the
{\it conditional} mass function in regions with the local background
density $\delta$, respectively, and $p_{dm}(\delta)$ is the PDF of the
dark matter density field.  
To evaluate the first term of the right
hand side of equation (\ref{eq:pMdelta2}), i.e. the excess number density of
halos, we adopt two approaches. The first is to directly use the mass
functions derived from the extended model of Press-Schechter (1974); 
see, for example, Bond et al.~(1991). Lacey \& Cole (1991) and 
Mo \& White (1996).  In practice, we adopt
the conditional mass function given by Sheth \& Tormen (2002) that
even takes into account the effect of the ellipsoidal collapse.
The other approach is to use a simple approximation,
$n(M|\delta)\simeq [1+b(M)\delta] n(M)$.  Note that this approximation
is valid only if the smoothing scale defining the local density 
is much larger than the halo size, or equivalently if the total mass
enclosed within the smoothing scale 
is much larger than the halo mass. 
Adopting this, equation (\ref{eq:pMdelta2}) reduces to
\begin{equation}
\label{eq:pMdelta}
p(M|\delta)=[1+b(M)\delta]p_{dm}(\delta).
\end{equation}
Mo \& White (1996) developed a Lagrangian halo bias model based on 
the extended Press-Schechter model. 
Jing (1998) and Sheth \& Tormen (1999) discussed the correction for 
the mass dependence of the halo bias and proposed modified fitting functions 
that give better fits to numerical simulation results.  
In what follows, we use the linear bias model of Sheth \& Tormen (1999).
Although the former approach may lead to a better model than 
the latter, we examine the latter approximation to see if the simple 
approach could provide an acceptable model.
Fig.~\ref{fig:condmf} compares the above two theoretical predictions 
(dashed and dotted lines) for $n(M|\delta)/n(M)$ with the $N$-body data
(solid histograms) for ranges of the local background density. 
The simple approximation $n(M|\delta)\simeq [1+b(M)\delta]
n(M)$ turns out to provide a good fit to the $N$-body data for smaller mass
ranges $M_{\rm halo}< 10^{13\sim 14}h^{-1}M_\odot$, but
underestimates the conditional probability for larger mass ranges.
The conditional mass function by Sheth \& Tormen (2002) gives good fits
for high local densities ($\delta>1$, upper two cases in the plots),
while fits become degraded for low densities ($\delta<1$, lower two 
cases in the plots).
Importantly, it reproduces the cut-off in the conditional mass
function at a large mass nicely in marked contrast to the approximation
which does not have the cut-off.

The dark matter PDF is known to be well fitted to the log-normal
distribution (e.g., Coles \& Jones 1991; Kofman et al. 1994; Kayo,
Taruya \& Suto 2001; see also Taruya, Hamana \& Kayo 2003 for a possible
explanation of the origin of the log-normal PDF)
\begin{equation}
\label{eq:lognormalPDF}
p_{ln}(\delta)={1 \over {(2\pi\sigma_1^2)^{1/2}}}
\exp\left\{-{{[\ln(1+\delta)+\sigma_1^2/2]^2} \over {2 \sigma_1^2}}\right\}
{1\over{1+\delta}}
\end{equation}
where $\sigma_1$ is related to the variance of the nonlinear density
field by
\begin{equation}
\label{eq:sigma1}
\sigma_1^2(R) = \ln[1+\sigma_{nl}^2(R)],
\end{equation}
with
\begin{equation}
\label{eq:sigmanl}
\sigma_{nl}^2(R) = {1 \over {2\pi^2}} \int dk~ k^2 P_{nl}(k) W^2(kR).
\end{equation}
In the above expression, $P_{nl}(k)$ denotes the nonlinear matter power
spectrum for which we compute using the fitting function of Peacock \&
Dodds (1996).

\section{Results}

Fig.~\ref{fig:sigmav_z} compares our model predictions with
$N$-body simulations (filled circles); long-dashed curves correspond to
equation (\ref{eq:pMdelta2}) with the conditional mass function by 
Sheth \& Tormen (2002), while solid curves correspond to the approximation
equation (\ref{eq:pMdelta}).  It is clear that our models (especially the 
former one) substantially improve the accuracy of the prediction compared 
with that of linear theory (dotted lines) except for a very high mass 
range of $M>10^{14}h^{-1}M_\odot$.  
However, the predictions of the model adopting the approximation 
equation (\ref{eq:pMdelta}) rise very rapidly at a high mass range
of $M>10^{14}h^{-1}M_\odot$ and significantly deviate from 
the simulation results.
This is explained as follows:
As shown in Fig.~\ref{fig:condmf}, the conditional mass function 
computed from $N$-body simulation declines at a high mass range.
However, the approximation equation (\ref{eq:pMdelta}) does not reproduce
this feature but keeps on rising toward a large halo mass.
Consequently, the approximated model over-estimates the contribution from
moderately over density regions.
Here, it must be noted that with our choice of the smoothing scale 
($R_{\rm local}= 18$, 15 and 13$h^{-1}$Mpc for $z=1$, 2 and 3, respectively), 
the most of the matter in the universe is in regions where the local 
density contrast being $-1<\delta<1$. 
Therefore, a discrepancy in the conditional mass function at a 
high density range of $\delta>1$ between the model and the simulation
does not have a large influence on the model prediction.
Adopting the conditional mass function proposed by Sheth \& Tormen (2002),
which reasonably reproduces the decline as shown in Fig.~\ref{fig:condmf}, 
improves the accuracy of the prediction very well.
In particular, this model predicts a flatter slope than that of the 
linear theory prediction, and it is very close to the simulation results.
We, however, note that the model prediction adopting Sheth \& Tormen 
conditional mass function rises at very large mass scales ($M>10^{15}$) 
and deviates from the simulation.
The reason of this is that the conditional mass 
function becomes less accurate at such mass range.
Because at that mass range, the number density of halos is too small to
have a meaningful statistic, this disagreement is not serious.

Although there is still room for improvement in the model, our model 
adopting the conditional mass function of Sheth \& Tormen (2002), 
provides a good fit to the $N$-body data over a wide range of the halo 
mass.
To improve the model prediction, it may be necessary to adopt
a further developed conditional mass function.
Developing an accurate conditional mass function is beyond the scope of 
the present paper, and hence we do not attempt this here.
In the following discussions, we adopt the model adopting Sheth \& Tormen
conditional mass function.

\begin{figure}
\begin{center}
\begin{minipage}{8.4cm}
\epsfxsize=8.4cm 
\epsffile{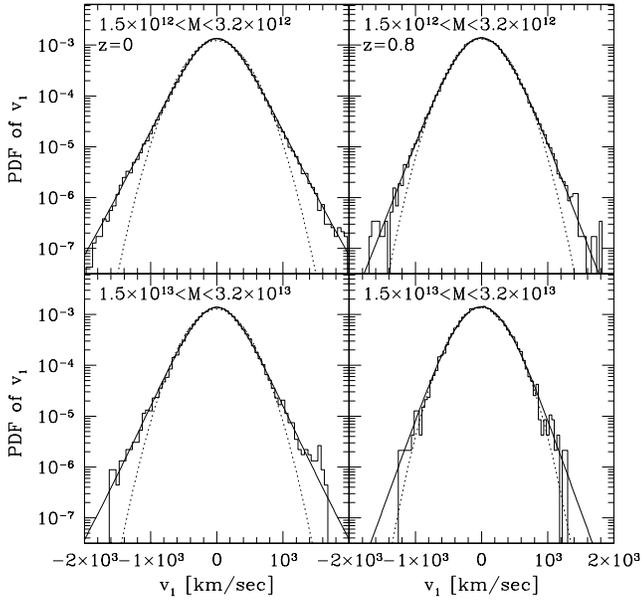}
\end{minipage}
\end{center}
\caption{PDFs of one-dimensional halo
 velocity $v_1$ at $z=0$ (left panels) and $z=0.8$ (right panels) and for 
limited mass ranges (denoted in each plot). 
Histograms show the measurement from the $N$-body simulations.
Solid curves show our model prediction, while dotted curves represent 
the Gaussian distribution with $\sigma$ computed from the $N$-body simulation.}
\label{fig:pdf_v1x4}
\end{figure}

As Sheth \& Diaferio (2001) originally showed, the model presented in
this paper is applied to model the PDF of halo peculiar velocities
\begin{equation}
\label{eq:PDFv}
P(v|M)=
{{\int d \delta~p(M|\delta) P(v|M,\delta)} 
\over 
{\int d \delta~p(M|\delta)}},
\end{equation}
where $P(v|M)$ is the PDF of peculiar velocities of halos with mass $M$,
and $P(v|M,\delta)$ is the PDF for halos at regions with the local
background density $\delta$.  The PDF of one-dimensional peculiar
velocities (which we denote by $v_1$), $P(v_1|M,\delta)$, is well
approximated by the Gaussian distribution (Sheth \& Diaferio 2001).  We
compute $\sigma$ of the Gaussian PDF using equation (\ref{eq:sigmahalo})
with a reasonable isotropic assumption $\langle v_1^2 \rangle =\langle
v^2 \rangle/3$.  The results are plotted in Fig.~\ref{fig:pdf_v1x4}
(solid lines), and are compared with $N$-body simulations (histograms)
and Gaussian PDF (dotted lines).  As Fig.~\ref{fig:pdf_v1x4} clearly
shows, our model agrees with the $N$-body simulation very well.
Also it is seen from the Figure that the PDF of one-dimensional peculiar 
velocity consists of two
parts; a Gaussian core and an exponential tail.  The exponential tail
arises from the sum of Gaussian distributions with different dispersions
(Sheth \& Diaferio 2001; Kuwabara, Taruya \& Suto 2002).  Our model 
nicely reproduces the deviation from the Gaussian distribution at
large peculiar velocities and agrees with the simulation very well.

\begin{figure}
\begin{center}
\begin{minipage}{8.4cm}
\epsfxsize=8.4cm 
\epsffile{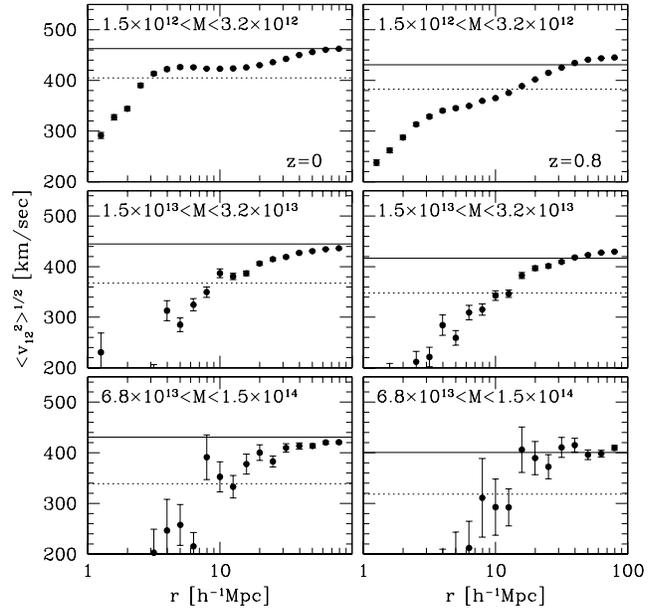}
\end{minipage}
\end{center}
\caption{Pairwise velocity dispersion of halos as a function of the
separation for limited mass ranges (denoted in each plot).  Filled
circles indicate the $N$-body results.  Dotted lines represent the
linear theory prediction for peaks neglecting velocity correlation,
$\langle v_{12}^2 \rangle^{1/2}=\sqrt{2/3}\sigma_p$, 
while solid lines show our model prediction 
$\langle v_{12}^2 \rangle^{1/2}=\sqrt{2/3}\sigma_{\rm halo}$.}
\label{fig:pairv}
\end{figure}

The model presented in this paper can be also applied to the pairwise
velocity dispersion of halos at large distances.  The model of the
pairwise velocity dispersion plays an important role in computing the
redshift distortion of the two-point correlation function of halos.
Because any velocity correlation between halos with large separations of
$r>20h^{-1}$Mpc may be safely ignored, the halo pairwise velocity
dispersion is given by $\langle v_{12}^2 \rangle = 2/3 \langle
v^2 \rangle$, where $v_{12}$ denotes the pairwise velocity of a pair of
halos.  Often $\sigma_p^2$ (equation \ref{eq:sigmap}) is used to compute
$\langle v^2 \rangle$ (e.g., Hamana et al.~2001); see Sheth et al.~2001a,b 
for an alternative approach.  However,
as shown in the previous sections, the linear theory prediction
underestimates the halo velocity and our model provides better fits to
$N$-body data.  We may, therefore, expect that our model improves
predictions of the pairwise velocity dispersion.  Fig.~\ref{fig:pairv}
compares the model prediction for the pairwise velocity dispersion
(solid lines) with the result of our $N$-body simulation (filled
circles). We also plot the linear theory prediction for peaks (dotted
lines).  As Fig.~\ref{fig:pairv} clearly shows, our model indeed
improves the accuracy of prediction at $r>20h^{-1}$Mpc.

\section{Summary and discussions}

We present a model of peculiar velocities of dark matter halos which
describes various statistical properties well and
significantly improves the accuracy of the simple linear theory
predictions, including the halo rms velocity dispersion, PDF, and the
pairwise velocity dispersion at large separations.  The model
predictions, however, becomes less accurate for very large halo masses
($M>10^{15}h^{-1}M_\odot$).
This is mainly due to the limitation of the adopted conditional halo mass
function, and may be reconciled by using an improved model.
Note that this shortcoming hardly causes any practical problem, because 
the number density of such very massive halos is too small to
have a meaningful statistic.

The model presented in this paper is characterized by a single free
parameter, $\mu(R_{\rm local})$, that we determine in an empirical
manner.  To do this, we need the appropriate smoothing scale for the
current purpose.  We make an anzats that $R_{\rm local}$ is in the
range between the linear and the non-linear scale and it should be given
by the relation $\sigma_0(R_{\rm local})=\sigma_{\rm local}$ for some
value in a range $0.1<\sigma_{\rm local}<1$.  This turns out to be
reasonably successful, and we have found that a choice of $\mu=0.6$ with
$\sigma_{\rm local}=0.5(1+z)^{-0.5}$ works well for all redshifts 
we consider in this paper ($z<2.2$). 
This choice gives the smoothing scales of $R_{\rm local}=18$, 15, 11, 8.5
and 5.6$h^{-1}$Mpc for $z=0$, 1, 3, 5 and 10, respectively.
However, we calibrated the model with $N$-body data at $0<z<2.2$ only, 
thus the validity of this procedure is 
questionable for very higher redshifts (although in practice velocities of
halos at very high redshifts are not observable currently).
A more sophisticated approach to
the choice of the smoothing scale is needed for a further improvement of
the model.  This may be closely connected to an understanding of the
role of the local environment in the structure formation, and is a
subject of work in progress.

\section*{Acknowledgments}

We thank the referee, Ravi Sheth, for detailed and constructive
comments on the earlier manuscript which significantly improves various
aspects of the paper.
We thank Joerg M. Colberg for alerting us to the mistake in the original 
version of this paper.
T.H.~acknowledges supports from Japan Society for
Promotion of Science (JSPS) Research Fellowships.  
Y.I.P. is supported in part by NKBRSF (G19990754) and by NSFC (No. 10125314).
This work was
supported in part by the Grant-in-Aid for Scientific Research of JSPS
(12640231).  Numerical computations presented in this paper were carried
out at the Astronomical Data Analysis Center of the National
Astronomical Observatory, Japan (project ID: mys02a).


\label{lastpage}

\begin{thebibliography}{99}

\bibitem{AGP01}
Aghanim N., G\'orski K. M. \& Puget J.-L., 2001, 374, 1

\bibitem{BGC94}
Bahcall N.A., Gramann M., Cen R. 1994, ApJ, 436, 23

\bibitem{BBKS86}
Bardeen J. M., Bond J. R., Kaiser N., Szalay A. S. 1986,
ApJ, 304, 15

\bibitem{Bondetal91}
Bond J. R., Cole S., Efstathiou G., Kaiser N.,
1991, ApJ, 379, 440

\bibitem{Borgani97}
Borgani S., da Costa L. N., Freudling W, Giovanelli R., Haynes
M. P., Salzer, J., Wegner, G. 1997, ApJ, 482, L121

\bibitem{Borgani00}
Borgani S., Bernardi M., da Costa L. N., Wegner G., Alonso M. V.,
Willmer C. N. A., Pellegrini P. S., Maia M. A. G., 2000, ApJ, 537, L1

\bibitem{Colberg00}
Colberg J. M., White S. D. M., MacFarland T. J., Jenkins A.,
Pearce F. R., Frenk, C. S., Thomas P. A., Couchman H. M. P., 2000,
MNRAS, 313, 229

\bibitem{Cole97}
Cole S., 1997, MNRAS, 386, 38

\bibitem{CJ1991} 
Coles P., Jones B. 1991, MNRAS, 248, 1

\bibitem{Colless01}
Colless M., Saglia R. P., Burstein D., Davies R. L., McMahan R. K.,
Wegner G., 2001, MNRAS, 321, 277

\bibitem{CE94}
Croft R. A. C., Efstathiou G., 1994, MNRAS, 268, L23

\bibitem{Dale97}
Dale D. A., Giovanelli R., Haynes M. P., Campusano L. E., Hardy E. 1999,
AJ, 118, 1489

\bibitem{DG96}
Diaferio A., Geller M., 1996, ApJ, 19

\bibitem{HT96}
Haehnelt M. G., Tegmark M., 1996, MNRAS, 279, 545

\bibitem{HYS02} 
Hamana T., Yoshida N., Suto Y., 2002, ApJ, 568, 455

\bibitem{HYSE01} 
Hamana T., Yoshida N., Suto Y., Evrard A. E., 2001, ApJ, 561, L143

\bibitem{Hudson99}
Hudson M. J., Smith R. J., Lucey J. R., Schlegel D. J., Davies R. L., 
1999, ApJ, 512, L79

\bibitem{Jenkins01} 
Jenkins A., Frenk C.S., White S.D.M., Colberg J.M., Cole S., Evrard 
A.E., Couchman H.M.P., Yoshida N., 2001, MNRAS, 321, 372

\bibitem{Jing98} 
Jing Y. P., 1998, ApJ, 503, L9

\bibitem{Jing00} 
Jing Y. P., 2000, ApJ, 535, 30

\bibitem{JS98} 
Jing Y. P., Suto Y., 1998, ApJ, 494, L5

\bibitem{JS00} 
Jing Y. P., Suto Y., 2000, ApJ, 529, L69

\bibitem{KAB00} 
Kashlinsky A., Atrio-Barandela F., 2000, ApJ, 536, L67

\bibitem{KTS01} 
Kayo  I., Taruya  A., Suto  Y., 2001, ApJ, 561, 22

\bibitem{KSS97} 
Kepner J. V., Summers F. J.,  Strauss M. A., 1997, New Astronomy, 2, 165

\bibitem{Kofmanetal94}
Kofman L., Bertschinger E., Gelb J.M. Nusser A., Dekel A,. 1994, ApJ,
420, 44

\bibitem{Kuwabara02}
Kuwabara T., Taruya A., Suto, Y. 2002, PASJ, 54, 503

\bibitem{LC93}
Lacey C., Cole S. 1993, MNRAS, 262, 627

\bibitem{LP94}
Lauer T. R., Postman M. 1994, ApJ, 425, 418

\bibitem{MW96}
Mo H. J., White S. D.M. 1996, MNRAS, 282, 347

\bibitem{Moscardini96}
Moscardini L., Branchini E., Brunozzi P. T., Borgani S., Plionis
M., Coles P. 1996, MNRAS, 282, 384

\bibitem{PD96} 
Peacock J.A., Dodds S.J., 1996, MNRAS, 280, L19

\bibitem{Peebles80}
Peebles, P. J. E., 1980, The Large-Scale Structure of the Universe 
Princeton Univ. Press, Princeton, NJ

\bibitem{PS74} 
Press W. H., Schechter P., 1974, ApJ, 187, 425

\bibitem{RL91}
Rephaeli Y., Lahav O. 1991, ApJ, 372, 21

\bibitem{Sheth01}
Sheth R. K., 1996, MNRAS, 279, 1310

\bibitem{SD01}
Sheth R. K., Diaferio A., 2001, MNRAS, 322, 901

\bibitem{SDHS01}
Sheth R. K., Diaferio A., Hui L., Scoccimarro R., 2001a, MNRAS, 326, 463

\bibitem{SHDS01}
Sheth R. K., Hui L., Diaferio A., Scoccimarro, R., 2001b, MNRAS, 325, 1288

\bibitem{ST99}
Sheth R. K., Tormen G., 1999, MNRAS, 308, 119

\bibitem{ST02}
Sheth R. K., Tormen G., 2002, MNRAS, 329, 61

\bibitem{SG99}
Suhhonenko I., Gramann M., 1999, MNRAS, 303, 77

\bibitem{SZ80}
Sunyaev R. A., Zel'dovich Y. B., 1980, MNRAS, 190, 413

\bibitem{thk02} 
Taruya A., Hamana T., I. Kayo, 2003, MNRAS, 339, 495

\bibitem{ts00} 
Taruya A., Suto Y. 2000, ApJ, 542, 559 

\bibitem{TB97}
Tormen G., Bertchinger, E., 1996,AJ, 472, 14

\bibitem{NSD01}
Yoshida N., Sheth R. K., Diaferio A., 2001, MNRAS, 328, 669

\end{thebibliography}
\end{document}